\documentclass[preprint,prd,aps,superscriptaddress,11pt]{revtex4-1}

\usepackage{graphicx}
\usepackage{amsmath}
\usepackage{amssymb}
\usepackage{dcolumn}
\usepackage{bm}
\usepackage{slashed}

\newcommand\meq{\overset{m=0}{=}}
\newcommand{\be}{\begin{equation}}
\newcommand{\ee}{\end{equation}}
\newcommand{\ba}{\begin{eqnarray}}
\newcommand{\ea}{\end{eqnarray}}
\newcommand{\no}{\nonumber \\}
\newcommand{\gsim}{\mathrel{\hbox{\rlap{\lower.55ex \hbox {$\sim$}}
                   \kern-.3em \raise.4ex \hbox{$>$}}}}
\newcommand{\lsim}{\mathrel{\hbox{\rlap{\lower.55ex \hbox {$\sim$}}
                   \kern-.3em \raise.4ex \hbox{$<$}}}}

\def\vq{{\vec q}}
\def\vp{{\vec p}}

\def\vj{{\vec j}}
\def\vE{{\vec E}}
\def\vB{{\vec B}}

\def\vo{{\vec\o}}

\def\roughly#1{\mathrel{\raise.3ex\hbox{$#1$\kern-.75em%
\lower1ex\hbox{$\sim$}}}}
\def\lsim{\roughly<}
\def\gsim{\roughly>}

\def\({\left(}
\def\){\right)}
\def\[{\left[}
\def\]{\right]}

\def\d{{\delta}}
\def\D{{\Delta}}
\def\o{{\omega}}
\def\O{{\Omega}}
\def\e{{\epsilon}}

\def\b{{\beta}}

\def\g{{\gamma}}

\def\p{{\pi}}

\def\m{{\mu}}
\def\n{{\nu}}
\def\r{{\rho}}
\def\s{{\sigma}}

\def\th{{\theta}}
\def\Th{{\Theta}}

\def\ps{{\psi}}

\def\na{{\nabla}}

\newcommand{\lag}{\langle}
\newcommand{\rag}{\rangle}

\newcommand{\pd}{{\partial}}

\date{\today}

\begin{document}

\title{\bf On Mass correction to Chiral Vortical Effect and Chiral Separation Effect}

\author{Shu Lin}
\email{linshu8@mail.sysu.edu.cn}
\affiliation{School of Physics and Astronomy, Sun Yat-Sen University, Zhuhai 519082, China}
\author{Lixin Yang}
\email{yanglx5@mail2.sysu.edu.cn}
\affiliation{School of Physics and Astronomy, Sun Yat-Sen University, Zhuhai 519082, China}

\begin{abstract}
We study coefficients of axial chiral vortical effect and chiral separation effect at finite temperature and vector chemical potential in massive theories. We present two independent methods of calculating the coefficients: one from field theory and the other using the mass term in axial anomaly equation. An ambiguity in the integration constant similar to hydrodynamic approach to axial chiral vortical effect exists in the latter, but can be fixed naturally in the presence of mass. We obtain perfect agreement between the methods. The results of axial chiral vortical effect and chiral separation effect indicate that the presence of mass generically suppresses the two coefficients, with less suppression at larger chemical potential. For phenomenologically relevant case of quark gluon plasma with three quark flavor, we find the correction is negligible.
\end{abstract}

\maketitle


\newpage

\section{Introduction}

It has been realized that axial anomaly can lead to macroscopic transport phenomena \cite{Vilenkin:1980fu,Erdmenger:2008rm,Banerjee:2008th,Son:2009tf,Neiman:2010zi,Landsteiner:2011cp}, including chiral magnetic effect \cite{Kharzeev:2004ey,Kharzeev:2007tn,Fukushima:2008xe}, chiral separation effect \cite{Metlitski:2005pr,Son:2004tq}, (vector and axial) chiral vortical effect \cite{Erdmenger:2008rm,Banerjee:2008th,Neiman:2010zi,Landsteiner:2011cp} etc. In the chiral limit, the corresponding transport coefficients can be derived either by hydrodynamics \cite{Son:2009tf,Neiman:2010zi} or by chiral kinetic theory \cite{Son:2012wh,Son:2012zy,Stephanov:2012ki,Pu:2010as,Chen:2012ca,Hidaka:2016yjf,Manuel:2013zaa,Manuel:2014dza,Huang:2018wdl}.
\begin{align}\label{chiral_full}
  &\vj=C\m_5e\vB+2C\m\m_5{\vo},\no
  &\vj_5=C\m e\vB+C\big[(\m^2+\m_5^2)+\frac{\pi^2T^2}{3}\big]\vo,
\end{align}
with $\vB$ and $\vo$ being magnetic field and fluid vorticity respectively. For quantum electrodynamics (QED), $C=\frac{1}{2\p^2}$. The first line of \eqref{chiral_full} corresponding to vector current contains chiral magnetic effect (CME) and vector chiral vortical effect (VCVE). Their existence rely on chiral imbalance, characterized by axial chemical potential $\m_5$. They are accessible in heavy ion collisions experiments \cite{Kharzeev:2007jp,Kharzeev:2010gr}, in which both strong magnetic field and vorticity can be produced in off-center heavy ion collisions \cite{Skokov:2009qp,Deng:2016gyh,Pang:2017bjc}. The second line of \eqref{chiral_full} corresponding to axial current contains chiral separation effect (CSE) and axial chiral vortical effect (ACVE), which are not directly accessible. Experimental search of CME and VCVE suffer from different background contamination from collective flow \cite{Voloshin:2010ut}, conservation law \cite{Schlichting:2010qia,Bzdak:2009fc} and resonance decays \cite{Wang:2009kd}. There have been continuous efforts in excluding these background experimentally \cite{Voloshin:2010ut,Wang:2012qs,Bloczynski:2013mca,Deng:2016knn,Wen:2016zic,Xu:2017qfs}. Phenomenological description of chiral transports has been carried out in the framework of anomalous hydrodynamics \cite{Kharzeev:2015znc,Huang:2015oca,Liao:2014ava} and stochastic anomalous hydrodynamics \cite{Lin:2018nxj}. In the latter, the axial charge is treated as a stochastic variable.

The anomalous hydrodynamics includes all the transports in \eqref{chiral_full}. Successful application of anomalous hydrodynamics requires precise knowledge about the transport coefficients. \eqref{chiral_full} has been derived by assuming axial charges is conserved, which is not true when we consider explicit axial symmetry breaking by fermion mass. Indeed, it is known at the discovery of CSE that mass correction exists for the corresponding transport coefficient \cite{Metlitski:2005pr,Gorbar:2013upa,Guo:2016dnm}. For ACVE, the situation is more interesting because it involves not only axial anomaly, but also gravitational anomaly \cite{Landsteiner:2011cp,Hou:2012xg,Golkar:2012kb}. The aim of this paper is to study mass correction to both ACVE and CSE coefficients. On one hand, it provides microscopic input for anomalous hydrodynamics for more realistic situation. On the other hand, the mass effect might shed some light on the role of gravitational anomaly.

The paper is organized as follows: in Section II, we give an intuitive derivation of ACVE and CSE coefficients from axial anomaly equation with mass term contribution. In Section III, we derive Kubo formula for ACVE and CSE in massive theories. In a restricted setting, the Kubo formula is formally the same as massless case. Section IV is devoted to the computation of ACVE and CSE coefficients for massive theories. The resulting coefficients are then compared to those obtained from the intuitive derivation. We find perfect agreement between them. We summarize and discuss possible extension of our work in Section V.

\section{Mass Correction and Pseudoscalar Condensate}

In this section, we present an intuitive derivation of ACVE and CSE coefficients based on axial anomaly equation for massive quantum electrodynamics (QED). For pedagogical purpose, we illustrate the derivation with massless QED first. The corresponding axial anomaly equation is given by
\begin{align}\label{anomaly}
\pd_\m j_5^\m=-\frac{e^2}{2\pi^2}E\cdot B.
\end{align}
Since ACVE requires the presence of fluid vorticity, we consider \eqref{anomaly} in a fluid cell with external fields $E$, $B$ and $\o$. Here $E$ and $B$ correspond to electric and magnetic fields in the local rest frame of fluid cell defined with fluid velocity $u^\m$ as $E^\m=F^{\m\n}u_\n$, $B^\m=\frac{1}{2}\e^{\m\n\r\s}u_\n F_{\r\s}$. The vorticity $\o^\m$ is defined as $\o^\m=\frac{1}{2}\e^{\m\n\r\s}u_\n \na_\r u_\s$. Throughout we use most minus signature.
To proceed, we specialize to static case. We can then express the electric field as gradient of chemical potential. This allows us to rewrite \eqref{anomaly} as
\begin{align}\label{rewrite}
  (2\p^2)\na\cdot\vj_5&=-e^2\vE\cdot\vB=\na\m \cdot e\vB \no
  &=\na\cdot(e\m\vB)-e\m\na\cdot\vB \no
  &=\na\cdot(e\m\vB)+2e\m(\vE\cdot\vo) \no
  &=\na\cdot(e\m\vB)+2\m(\na\mu\cdot\vo) \no
  &=\na\cdot(e\m\vB)+\na\cdot(\mu^2\vo).
\end{align}
In the above, we have used the relation $\na\cdot\vo=0$, $\na\cdot\vB=2\vE\cdot\vo$, which are static limits of the covariant relations $\pd_\m\o^\m=0$, $\pd_\m B^\m=2\o\cdot E$ \cite{Chen:2012ca}. It follows from \eqref{rewrite} that
\begin{align}\label{qed_j5}
  \vj_5=\frac{1}{2\p^2}\m e\vB+\frac{1}{2\p^2}\m^2\vo.
\end{align}
Note that \eqref{qed_j5} reproduces CSE and $\m$-dependent part of ACVE in massless QED. It is possible to include an additional term $\na\cdot(\m^2\vo)\to\na\cdot(\m^2\vo+\#T^2\vo)$. This would give rise to a $\m$-independent contribution to ACVE coefficient. However the coefficient is not constrained in this derivation, similar to the ambiguity in the integration constant found early in hydrodynamic approach to anomalous transports \cite{Neiman:2010zi}. It is attributed to gravitational anomaly contribution \cite{Landsteiner:2011cp}. Below we will see that in massive theory there is a natural choice of additional term that allows us to determine contributions to ACVE from both axial anomaly and gravitational anomaly.

Now we generalize the above derivation to massive QED.
It has been found in \cite{Fang:2016uds} that the leading electromagnetic interaction correction to axial anomaly equation \eqref{anomaly} is given by:
\begin{align}\label{wang}
  \pd_\m j_5^\m=-\frac{1}{2\p^2}(e^2E\cdot B)C_1(m,\b,\m)-\frac{m^2}{2\p^2}\b(eE\cdot\o)C_2(m,\b,\m).
\end{align}
The coefficient functions $C_1$ and $C_2$ are defined as
\begin{align}\label{nonconservation_c1c2}
  &C_1(m,\b,\m)=\b\int_0^{\infty}dq\Big[\frac{e^{\b(E_q-\m)}}{\big(e^{\b(E_q-\m)}+1\big)^2}+\frac{e^{\b(E_q+\m)}}{\big(e^{\b(E_q+\m)}+1\big)^2}\Big],\no
  &C_2(m,\b,\m)=\int_0^{\infty}dq\frac{1}{E_q}\Big[\frac{e^{\b(E_q-\m)}}{\big(e^{\b(E_q-\m)}+1\big)^2}-\frac{e^{\b(E_q+\m)}}{\big(e^{\b(E_q+\m)}+1\big)^2}\Big],
\end{align}
with $E_q=\sqrt{q^2+m^2}$.
By repeating the same procedure as above, and noting the $\m$ dependence of $C_1$ and $C_2$, we obtain
\begin{align}\label{rewrite_massive}
  (2\p^2)\na\cdot\vj_5=\na\cdot(F_1(\b,m,\m)e\vB)+2\na\cdot(F_3(\b,m,\m)\vo)+m^2\na\cdot(F_2(\b,m,\m)\vo).
\end{align}
The coefficient functions are defined as
\begin{align}\label{chiral_current_f123}
  &F_1(\b,m,\m)=\int^\m d\m'C_1(\b,m,\m')=\int_0^{\infty}dq\tilde f_-(E_q)+a_1(\b,m),\no
  &F_2(\b,m,\m)=\int^\m d\m'C_2(\b,m,\m')=\int_0^{\infty}dq\frac{1}{E_q}\tilde f_+(E_q)+a_2(\b,m),\no
  &F_3(\b,m,\m)=\int^\m d\m'F_1(\b,m,\m')=\frac{1}{\b}\int_0^{\infty}dq L(E_q)+a_3(\b,m),
\end{align}
with the functions $\tilde f_\pm(E_q)$ and $L(E_q)$ defined in terms of Fermi-Dirac distribution function $\tilde n(x)=\frac{1}{e^{\b x}+1}$ as $\tilde f_{\pm}(E_q)=\tilde n(E_q-\m)\pm\tilde n(E_q+\m)$ , $L(E_q)=-\ln\big(\tilde n(-E_q+\m)\big)-\ln\big(\tilde n(-E_q-\m)\big)$.
The integration constants $a_n(\b,m)(n=1,2,3)$ are fixed as follows: $a_1=0$ because there is no C-odd function depending on $\b$ and $m$ only. $a_2$ and $a_3$ are not constrained by C-symmetry, but the most ``natural choice'' is to require that they can be expressed as integration of distribution function, but this is not possible because they are $\m$ independent. This fixes $a_2=a_3=0$.
Consequently, the ACVE and CSE in massive case readily follow from \eqref{rewrite_massive} and \eqref{chiral_current_f123}.
\begin{align}\label{constitutive_relations}
  \vj_5=\overline\s_V\vo+\overline\s_B\vB,
\end{align}
with
\begin{align}\label{CSE_ACVE_coefficients}
  &\overline\s_V=\frac{1}{2\p^2}\Big(2F_3(\b,m,\m)+m^2F_2(\b,m,\m)\Big),\no
  &\overline\s_B=\frac{e}{2\p^2}F_1(\b,m,\m),
\end{align}

\section{Kubo formula for massive theory}

Next we derive Kubo formula needed for field theoretic computation of ACVE and CSE coefficients.
The Kubo formulas for anomalous transport coefficients have been derived in \cite{Amado:2011zx} assuming axial current is conserved. For massive theory, axial current is not conserved due to the mass term. However as we will show the Kubo formula for CSE and ACVE remain unchanged in a restricted setting. For the purpose of deriving Kubo formula, we turn on static but spatially inhomogeneous gauge fields $A_i$ and metric perturbation $h_{0i}$. They give rise to magnetic field and fluid vorticity:
\begin{align}\label{b_omega}
  &B^i=-\e^{ijk}\pd_j A_k, \no
  &\o^i=-\frac{1}{2}\e^{ijk}\na_j u_k=-\frac{1}{2}\e^{ijk}\pd_jh_{0k}.
\end{align}
In arriving at $\o^i$, we implicitly choose a frame, in which the fluid velocity remains unchanged i.e $u^\m=(1,0,0,0)$ in the presence of the external fields. Since $E$ is absent, $B$ and $\o$ cannot induce pseudoscalar condensate. This can be seen from analysis of discrete symmetries: The mass term is odd in parity (P) and time reversal (T), and even in charge conjugation (C). On the other hand, $B$ is T-even, P-odd and C-odd; $\o$ is T-even, P-odd and C-even. The only possibilities to form quantities with the same symmetry properties as the mass term are: $\na\m\cdot\vB$, $\m\na\m\cdot\vo$ and $T\na T\cdot\vo$. Throughout this paper, we consider state with constant $\m$ and $T$. This excludes the possibility of generating a nonvanishing mass term, allowing us to treat axial charge as a conserved quantity.
It follows that we can use the known constitutive equation for axial current \cite{Son:2009tf}:
\begin{align}\label{consti}
  j_5^\m(x)=n_5(x)u^\m(x)+\s_BB^\m(x)+\s_V\o^\m(x)+\cdots,
\end{align}
where $\s_B$ and $\s_V$ are CSE and ACVE coefficients. The terms omitted are not relevant to us. 
Plugging \eqref{b_omega} into \eqref{consti}, we obtain for the spatial components
\begin{align}\label{ji}
  j_5^i(x)=-\s_B\e^{ijk}\pd_j A_k(x)-\frac{1}{2}\s_V\e^{ijk}\pd_jh_{0k}(x).
\end{align}
Fourier transform of \eqref{ji} gives
\begin{align}\label{ji_Fourier}
  j_5^i(p)=-\s_B\e^{ijk}ip_j A_k(p)-\frac{1}{2}\s_V\e^{ijk}ip_jh_{0k}(p).
\end{align}
Upon permutation of indices, we obtain the following Kubo formula for CSE and ACVE coefficients
\begin{align}\label{kubo}
  &\s_B=\lim_{p_k\to0}\frac{-1}{2ip_k}\e^{ijk}G^R_{ij}(p)\lvert_{p_0=0},\no
  &\s_V=\lim_{p_k\to0}\frac{-1}{ip_k}\e^{ijk}G^R_{i,0j}(p)\lvert_{p_0=0},
\end{align}
where the retarded correlators are defined as
\begin{align}
  &G^R_{ij}(p)=i\int d^4xe^{ip\cdot x}\lag\big[ j_5^i(x),j^j(0)\big]\rag\th(x^0), \no
  &G^R_{i,0j}(p)=i\int d^4xe^{ip\cdot x}\lag\big[ j_5^i(x),T^{0j}(0)\big]\rag\th(x^0).
\end{align}

\section{Axial Chiral Vortical Effect and Chiral Separation Effect}

In this section, we show the computation of the ACVE coefficient in massive QED in detail and sketch the computation of CSE coefficient since the latter involves the same procedure and technics as the former.

\subsection{ACVE coefficient}
The operators in $G^R_{i,0j}$ are
\begin{align}\label{currents_dirac}
  &j_5^i=\bar\ps\g^i\g_5\ps,\no
  &T^{0i}=\frac{i}{2}\bar\ps(\g^0\pd^i+\g^i\pd^0)\ps.
\end{align}
$j_5^i$ is standard. We give a derivation of $T^{0i}$ in massive QED in appendix.
For simplicity, we consider state with finite $\m$ and $T$ with $\m_5=0$. We work in imaginary-time formalism. The relevant massive fermion propagator is given by
\begin{align}\label{propagator}
  S(Q)=\frac{1}{\g^0(i\tilde{\o}_m+\m)-\vec\g\cdot\vq-m}=\frac{(i\tilde{\o}_m+\m+m\g^0)\g^0-\vec\g\cdot\vq}{(i\tilde{\o}_m+\m)^2-q^2-m^2}\equiv\frac{A(Q)}{D(Q)},
\end{align}
with $q=\lvert\vq\lvert$ and $\vec\g=(\g_1,\g_2,\g_3)$. The last equality of \eqref{propagator} defines $A(Q)$ and $D(Q)$ as numerator and denominator respectively.
Using $E_q=\sqrt{q^2+m^2}$ and 
$\frac{1}{\D_{\pm}(Q)}=\frac{1}{i\tilde{\o}_m+\m\mp E_q}$, we can write
\begin{align}\label{split_d}
  \frac{1}{D(Q)}=[\D_+(Q)-\D_-(Q)]\frac{1}{2E_q}=[\D_+(Q)+\D_-(Q)]\frac{1}{2(i\tilde{\o}_m+\m)}.
\end{align}
The components we need for ACVE coefficient is given by
\begin{align}\label{correlator_acve}
  G_V(P)&\equiv-\e^{ijk}G^E_{i,0j}(P)=\frac{1}{2\b}\sum_{\tilde{\o}_m}\int\frac{d^3q}{(2\p)^3}\epsilon_{ijk}Tr[S(Q)\g^i\g_5S(P+Q)(\g^0q^j+\g^ji\tilde{\o}_m)]\no
  &=G_V^{0j}(P)+G_V^{j0}(P),
\end{align}
with $G_V^{0j}$ and $G_V^{j0}$ corresponding to the terms $\g^0q^j$ and $\g^ji\tilde\o_m$ in the bracket respectively. The Euclidean correlator $G^E(P)$ is related to the retarded correlator by analytic continuation $G^R(P)=G^E(P)\lvert_{i\o_n\to p_0+i\varepsilon}$.

We first calculate $G_V^{0j}$. Using \eqref{propagator} and \eqref{split_d}, we have 
\begin{align}\label{correlator_0j}
  G_V^{0j}(P)&=\frac{1}{8\b}\sum_{\tilde{\o}_m}\int\frac{d^3q}{(2\p)^3}\epsilon_{ijk}Tr[A(Q)\g^i\g_5A(P+Q)\g^0q^j]\no
  &\quad\times\frac{1}{E_qE_{p+q}}\sum_{u,v=\pm}uv\D_u(Q)\D_v(P+Q).
\end{align}
The Dirac trace in the integrand of \eqref{correlator_0j} can be evaluated as
\begin{align}\label{trace_0j}
\epsilon_{ijk}Tr[\g_{\m}\g^i\g_5\g_{\n}\g^0]a^{\m}b^{\n}=4i(a_jb_k-a_kb_j).
\end{align}
Using \eqref{correlator_0j} and \eqref{trace_0j} we have
\begin{align}
  G_V^{0j}(P)&=\frac{i}{2\b}\sum_{\tilde{\o}_m}\int\frac{d^3q}{(2\p)^3}q^j(q_jp_k-q_kp_j)\frac{1}{E_qE_{p+q}}\sum_{u,v=\pm}uv\D_u(Q)\D_v(P+Q).
\end{align}
Both $\tilde{\o}_m=(2m+1)\p T$ and $\tilde{\o}_m+\o_n$ are fermionic, because $\o_n=2n\p T$ is bosonic. The sum over fermionic Matsubara frequencies gives
\begin{align}\label{frequency_sum_0j}
\frac{1}{\b}\sum_{\tilde{\o}_m}\D_u(Q)\D_v(P+Q)=\frac{u\,\tilde n(E_q-u\m)-v\,\tilde n(E_{p+q}-v\m)+\frac{1}{2}(v-u)}{i\o_n+uE_q-vE_{p+q}}.
\end{align}
The analytic continuation, i.e. replacing $i\o_n$ by $p_0+i\varepsilon$ in \eqref{frequency_sum_0j}, gives the retarded correlator
\begin{align}
  G_V^{0j}(P)&=-\frac{i}{2}\int\frac{d^3q}{(2\p)^3}(q^2p_k-\vq\cdot\vp q_k)\frac{1}{E_qE_{p+q}}\no
  &\quad\times\sum_{u,v=\pm}\frac{v\,\tilde n(E_q-u\m)-u\,\tilde n(E_{p+q}-v\m)+\frac{1}{2}(u-v)}{p_0+i\varepsilon+uE_q-vE_{p+q}}.
\end{align}
Here the term proportional to $\frac{1}{2}(u-v)$ corresponds to vacuum contribution. Since we are mainly interested in medium effect, we do not keep it. Applying the relabeling $\vq\to-\vq-\vp$ and $u\leftrightarrow -v$ in the part involving $u\,\tilde n(E_{p+q}-v\m)$, we get
\begin{align}
  G_V^{0j}(P)&=-\frac{ip_k}{2}\int\frac{d^3q}{(2\p)^3}\Big(q^2-\frac{(\vq\cdot\vp)^2}{p^2}\Big)\frac{1}{E_qE_{p+q}}\sum_{u,v=\pm}v\,\frac{\tilde n(E_q-\m)+\tilde n(E_q+\m)}{p_0+i\varepsilon+uE_q-vE_{p+q}}.
\end{align}
Here we have replaced $q_k$ by $p_k(\vq\cdot\vp)/p^2$ by rotational symmetry of $G_V^{0j}$. By summing over $v$ and integrating over angles, we get
\begin{align}\label{final_0j}
  G_V^{0j}(P)&=\frac{ip_k}{4\p^2}\int_0^{\infty}q^2dq\tilde f_+(E_q)\frac{q^2}{E_q}\sum_{u=\pm}\frac{1}{(2pq)^3}\Big[\Big((2pq)^2-\big(\O_u^2-(p^2+q^2+m^2)\big)^2\Big)\no
  &\quad\times\ln\frac{\O_u^2-(p+q)^2-m^2}{\O_u^2-(p-q)^2-m^2}-2\big(\O_u^2-(p^2+q^2+m^2)\big)2pq\Big],
\end{align}
where $\O_u=p_0+i\varepsilon+uE_q$.

Similarly, for $G_V^{j0}$, the trace identity is evaluated as
\begin{align}\label{trace_j0}
  \epsilon_{ijk}Tr[\g_{\m}\g^i\g_5\g_{\n}\g^j]a^{\m}b^{\n}=8i(a^0b^k-a^kb^0)=8i(a_kb_0-a_0b_k).
\end{align}
Using \eqref{propagator}, \eqref{split_d}, \eqref{correlator_acve} and \eqref{trace_j0}, we get
\begin{align}
  G_V^{j0}(P)&=\frac{4i}{\b}\sum_{\tilde{\o}_m}\int\frac{d^3q}{(2\p)^3}\Big(q_k(i\tilde{\o}_m+\o_n+\m)-(i\tilde{\o}_m+\m)(p_k+q_k)\Big)i\tilde{\o}_m\frac{1}{D(Q)D(P+Q)}\no
  &=\frac{i}{\b}\sum_{\tilde{\o}_m}\int\frac{d^3q}{(2\p)^3}\Big(u\frac{q_k}{E_q}-v\frac{p_k+q_k}{E_{p+q}}\Big)i\tilde{\o}_m\sum_{u,v=\pm}\D_u(Q)\D_v(P+Q).
\end{align}
The frequency sum involving an extra $i\tilde{\o}_m$ can be written as
\begin{align}\label{frequency_sum_j0}
  &\frac{i\tilde{\o}_m}{\b}\sum_{\tilde{\o}_m}\frac{1}{i\tilde{\o}_m+\m-u\,E_q}\frac{1}{i\tilde{\o}_m+i\o_n+\m-v\,E_{p+q}}\\
  &=\frac{1}{\b}\sum_{\tilde{\o}_m}\Big(\frac{u\,E_q-\m}{i\tilde{\o}_m+\m-u\,E_q}-\frac{v\,E_{p+q}-\m-i\o_n}{i\tilde{\o}_m+i\o_n+\m-v\,E_{p+q}}\Big)\frac{1}{u\,E_q-v\,E_{p+q}+i\o_n}\no
  &=\frac{(u\,E_q-\m)\tilde n(\m-uE_q)-(v\,E_{p+q}-\m-i\o_n)\tilde n(i\o_n+\m-vE_{p+q})}{i\o_n+uE_q-vE_{p+q}}\no
  &=\frac{(u\,E_q-\m)\big(\frac{1+u}{2}-u\tilde n(E_q-u\m)\big)-(v\,E_{p+q}-\m-i\o_n)\big(\frac{1+v}{2}-v\tilde n(E_{p+q}-vi\o_n-v\m)\big)}{i\o_n+uE_q-vE_{p+q}}.\nonumber
\end{align}
We can neglect the term proportional to $i\o_n$ in the numerator because in the end $i\o_n\to p_0+i\varepsilon\to0$. The remaining $\tilde n(x)$ independent terms corresponding to vacuum contribution are not kept for the same reasons as before. By relabeling $\vq\to-\vq-\vp$ and $u\leftrightarrow-v$ in the part involving $(v\m-E_{p+q})\tilde n(E_{p+q}-v\m)$, one gets
\begin{align}\label{final_j0}
  G_V^{j0}(P)&=i\int\frac{d^3q}{(2\p)^3}\sum_{u,v=\pm}\frac{u\m\tilde f_-(E_q)-E_q\tilde f_+(E_q)}{\O_u-vE_{p+q}}\Big(u\frac{q_k}{E_q}-v\frac{p_k+q_k}{E_{p+q}}\Big)\no
  &=\frac{ip_k}{2\p^2}\int_0^{\infty}q^2dq\sum_{u=\pm}(E_q\tilde f_+(E_q)-u\m\tilde f_-(E_q))\Big[\frac{1}{2pq}\ln\frac{\O_u^2-(p+q)^2-m^2}{\O_u^2-(p-q)^2-m^2}-\no
  &\quad u\frac{qp_0}{E_qp(2pq)^2}\Big(4pq-\big(\O_u^2-(p^2+q^2+m^2)\big)\ln\frac{\O_u^2-(p+q)^2-m^2}{\O_u^2-(p-q)^2-m^2}\Big)\Big].
\end{align}
The result of $G_V$ is the sum of \eqref{final_0j} and \eqref{final_j0}, according to \eqref{correlator_acve}. Using
\begin{align}\label{zero_limit}
  &\lim_{p\to0}\lim_{p_0\to0}\sum_{u=\pm}\ln\frac{\O_u^2-(p+q)^2-m^2}{\O_u^2-(p-q)^2-m^2}=\frac{2p}{q},\no
  &\lim_{p\to0}\lim_{p_0\to0}\sum_{u=\pm}u\ln\frac{\O_u^2-(p+q)^2-m^2}{\O_u^2-(p-q)^2-m^2}=0,
\end{align}
we obtain the ACVE coefficient from \eqref{kubo} as
\begin{align}\label{coefficient_acve}
  \s_V=\frac{1}{2\p^2}\int_0^{\infty}\tilde f_+(E_q)\frac{2q^2+m^2}{E_q}dq\meq\frac{\m^2}{2\p^2}+\frac{T^2}{6}.
\end{align}
We verify in the last equality that it is consistent with \eqref{chiral_full} in the massless limit.

To compare $\s_V$ with ${\bar \s}_V$ in \eqref{CSE_ACVE_coefficients}, we first note that the term proportional to $m^2F_2$ in \eqref{CSE_ACVE_coefficients} is identical to the $m^2$ term in \eqref{coefficient_acve}. The remaining terms can be shown to be identical using integration by part:
\begin{align}\label{L-term}
  \int_0^\infty dq\frac{2q^2}{E_q}{\tilde f}_+(E_q)=-\int_0^\infty dq \frac{2q}{\b}\frac{\pd}{\pd q}L(E_q)=\int_0^\infty dq \frac{2}{\b}L(E_q)=2F_3(\b,m,\m).
\end{align}
Therefore, our ``natural choice'' of integration constants leads to the correct ACVE coefficient. Note that this was not possible before the introduction of fermion mass.

\subsection{CSE coefficient}
We start with components of Euclidean correlator needed for CSE coefficient:
\begin{align}\label{correlator_cse}
  G_B(P)&\equiv-\frac{1}{2}\e^{ijk}G^E_{ij}(p)=\frac{e}{2\b}\sum_{\tilde{\o}_m}\int\frac{d^3q}{(2\p)^3}\epsilon_{ijk}Tr[S(Q)\g^i\g_5S(P+Q)\g^j]
\end{align}
The computation of the Dirac trace is the same as \eqref{trace_0j} and the frequency sum is a repetition of the case of \eqref{frequency_sum_0j}, so we skip the details. After the analytic continuation, the result is
\begin{align}\label{analytic_cse}
  G_B(P)&=ie\int\frac{d^3q}{(2\p)^3}\sum_{u,v=\pm}\frac{u\,\tilde n(E_q-u\m)-v\,\tilde n(E_{p+q}-v\m)+\frac{1}{2}(v-u)}{p_0+i\varepsilon+uE_q-vE_{p+q}}\big(u\frac{q_k}{E_q}-v\frac{p_k+q_k}{E_{p+q}}\big).
\end{align}
By dropping the vacuum contribution proportional to $\frac{1}{2}(v-u)$ and relabeling $\vq\to-\vq-\vp$ and $u\leftrightarrow-v$ in the part involving $v\tilde n(E_{p+q}-v\m)$, we have
\begin{align}\label{final_cse}
  G_B(P)&=iep_k\int\frac{d^3q}{(2\pi)^3}\sum_{u,v=\pm}\frac{\tilde n(E_q-\m)-\tilde n(E_q+\m)}{p_0+i\varepsilon+uE_q-vE_{p+q}}\Big(u\frac{\vq\cdot\vp}{E_qp^2}-v\frac{p^2+\vq\cdot\vp}{E_{p+q}p^2}\Big)\no
  &=\frac{iep_k}{2\p^2}\int_0^{\infty}q^2dq\tilde f_-(E_q)\sum_{u=\pm}\frac{1}{4E_qp^3q}\Big[4pp_0q+\no
  &\quad\Big(2E_qp^2+p_0\big(\O_u^2-(p^2+q^2+m^2)\big)\Big)\ln\frac{\O_u^2-(p+q)^2-m^2}{\O_u^2-(p-q)^2-m^2}\Big].
\end{align}
Taking the limit $p\to0$ after $p_0\to0$, we obtain the CSE coefficient from \eqref{kubo} as
\begin{align}\label{coefficient_cse}
  \s_B=\frac{e}{2\p^2}\int_0^{\infty}dq\tilde f_-(E_q)\meq\frac{e\m}{2\p^2}.
\end{align}
We verify that it is consistent with \eqref{chiral_full} in the massless limit.
It is easy to see \eqref{coefficient_cse} is the same as $\overline\s_B$ in \eqref{CSE_ACVE_coefficients}.

\subsection{Discussion}

\begin{figure}
  \includegraphics[width=0.6\textwidth]{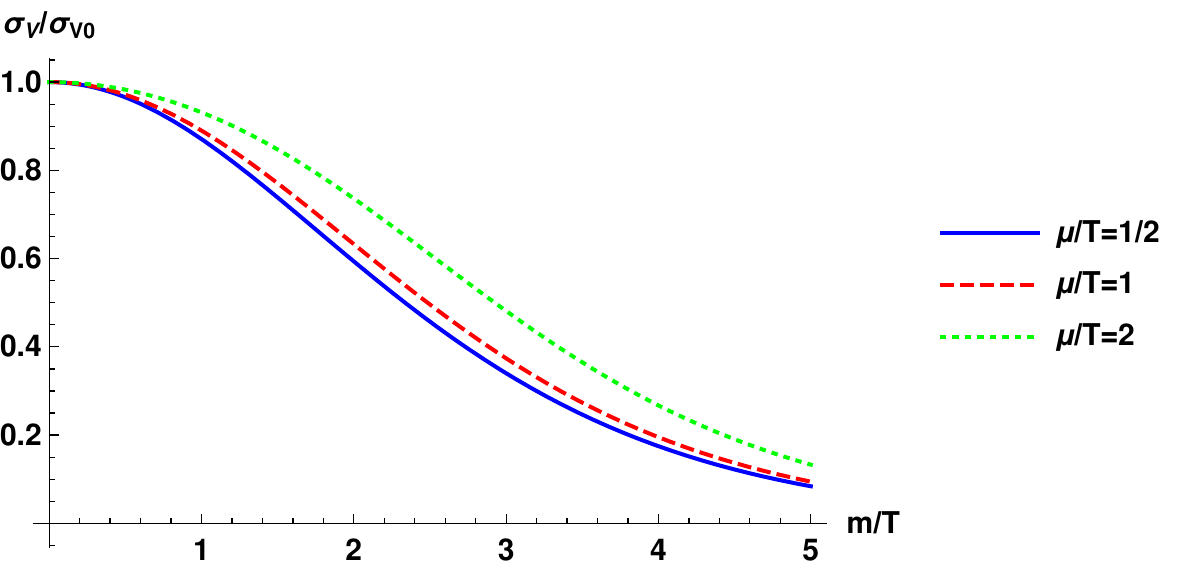}
  \caption{Mass dependence of $\s_V/s_{V0}$ at $\m/T=1/2$ (blue solid), $\m/T=1$ (red dashed) and $\m/T=2$ (green dotted). $\s_{V0}$ is the ACVE coefficient in massless limit \eqref{chiral_full}. The presence of mass generically suppress ACVE coefficient with less suppression at larger $\m$.}
  \label{fig:acve}
\end{figure}

\begin{figure}
  \includegraphics[width=0.6\textwidth]{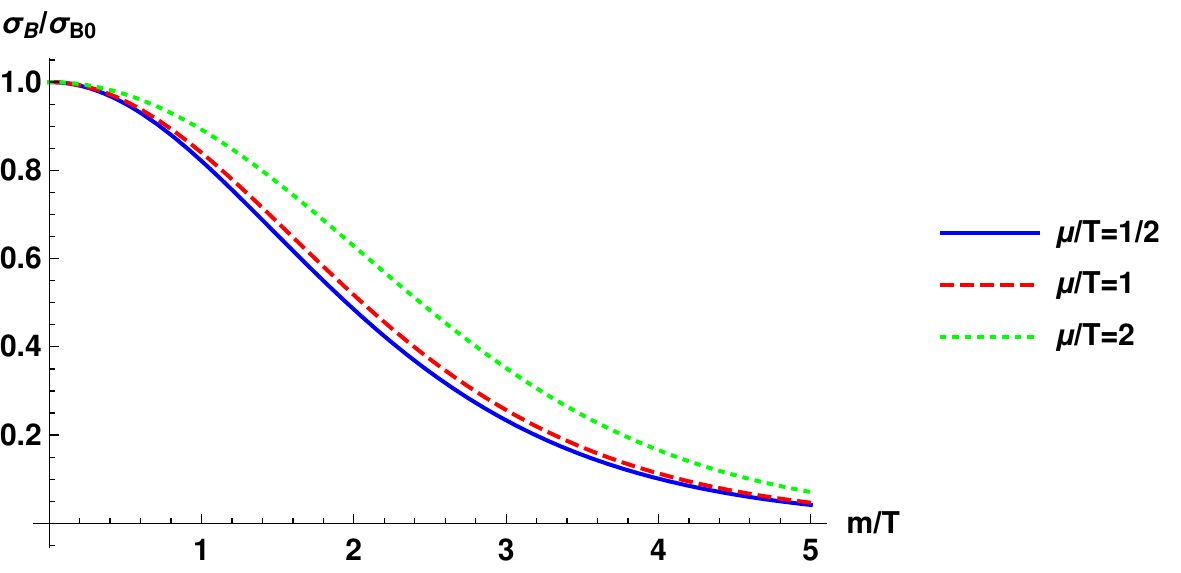}
  \caption{Mass dependence of $\s_B/s_{B0}$ at $\m/T=1/2$ (blue solid), $\m/T=1$ (red dashed) and $\m/T=2$ (green dotted). $\s_{B0}$ is the CSE coefficient in massless limit \eqref{chiral_full}. The presence of mass generically suppress CSE coefficient with less suppression at larger $\m$.}
  \label{fig:cse}
\end{figure}

We consider different limits of the coefficients $\s_V$ and $\s_B$. Let us begin with $\s_B$: at $T=0$, \eqref{coefficient_cse} can be easily evaluated to give $\s_B=\frac{1}{2\pi^2}\sqrt{\m^2-m^2}$, which is in agreement with the free limit of \cite{Gorbar:2013upa}. At $T\ne0$, $\s_B$ adopts the following small $m$ expansion:
\begin{align}\label{sigmaB_m}
  \s_B
  &=\frac{e}{2\p^2}\Bigg[\m+\frac{m^2}{2T}\frac{\pd}{\pd s}\Big(\textrm{Li}_s(-e^{\m/T})-\textrm{Li}_s(-e^{-\m/T})\Big)\Big\lvert_{s=-1}+O(m^4)\Bigg],
\end{align}
where $\textrm{Li}_s(z)$ is polylogarithm function. The coefficient of $m^2$ is a negative, non-monotonic function of $\m$ with the minimum at $\m/T\sim O(1)$. Interestingly, we find possible correction proportional to $m^2\ln m$ vanishes in the final result.
Turning to $\s_V$, we find its zero temperature limit can be obtained as $\s_V=\frac{1}{2\pi^2}\m\sqrt{\m^2-m^2}$. In this limit, the contribution related to gravitational anomaly is absent. Unlike $\s_B$, whose $\m=0$ limit vanishes by C-symmetry, $\s_V$ does have a non-vanishing $\m=0$ limit, whose small $m$ expansion can be obtained analytically as
\begin{align}
  \s_V
  &=\frac{1}{2\p^2}\Bigg[\m^2+\frac{\p^2T^2}{3}+m^2\Bigg(\frac{1}{2}+2\frac{\pd}{\pd s}\Big(\textrm{Li}_s(-e^{\m/T})+\textrm{Li}_s(-e^{-\m/T})\Big)\Big\lvert_{s=0}\Bigg)+O(m^4)\Bigg].
\end{align}
The coefficient of $m^2$ is a negative, monotonically decreasing function of $\m$. Surprisingly, we find possible correction proportional to $m^2\ln m$ also vanishes in the final result.
For more general parameters, we plot the $m$ dependence of ACVE and CSE coefficients in Fig.~\ref{fig:acve} and Fig.~\ref{fig:cse} respectively. We find the presence of mass generically suppress both ACVE and CSE coefficients, with less suppression at larger $\m$. At small $m$, we find the correction given by $m^2$. At large $m$, ACVE and CSE coefficients are expected to be exponentially suppressed.

The above results easily generalize to quantum chromodynamics (QCD), in which the ACVE and CSE coefficients are given by:
\begin{align}\label{qcd_coeff}
  &\s_V=\sum_f\frac{N_c}{2\p^2}\int_0^{\infty}dq\tilde f_+(\sqrt{q^2+m_f^2})\frac{2q^2+m_f^2}{E_q},\no
  &\s_B=\sum_f\frac{eN_c}{2\p^2}\int_0^{\infty}dq\tilde f_-(\sqrt{q^2+m_f^2}).
\end{align}
The only difference is an overall factor of $N_c$ and summation over quark flavors. It is interesting to see how mass corrected $\s_V$ and $\s_B$ affect phenomenology of heavy ion collisions. Taking three flavors of quark with $m_u=m_d\simeq 0$, $m_s=100{\rm MeV}$, we find mass correction to both $\s_B$ and $\s_V$ are within $1\%$ in the parameter range $50{\rm MeV}<\m<200{\rm MeV}$ and $200{\rm MeV}<T<400{\rm MeV}$. This justifies the use of $\s_V$ in chiral limit.

\section{Summary and outlook}

In this work, we compute ACVE and CSE coefficients in massive QED and QCD in the free limit. The results we obtain are in perfect agreement with an intuitive derivation based on mass correction to anomaly equation. Although ambiguity exists in the determination of ACVE coefficient in massless limit, we find that there is a ``natural choice'' in massive case that allows us to fix the ambiguity. The ``natural choice'' is based on the assumptions that the coefficients can be expressed as an integration of distribution function, which is well defined in free theory limit. It would be interesting to see if similar choice still exists in weakly interacting theory.

In massless theories, the ACVE coefficient contains contributions from axial and gravitational anomaly, which separates into terms proportional $\m^2$ and $T^2$. In massive theories, the separation is less obvious. A natural separation is to assign the $\m$-independent part $\s_V(\m=0)$ to gravitational anomaly, and the $\m$-dependent part $\s_V(\m)-\s_V(\m=0)$ to axial anomaly. Surprisingly, our choice of integration constants fix them simultaneously. Note that the intuitive derivation implicitly assumes a gradient expansion. However it is known that the gradient expansion breaks down for the contribution from gravitational anomaly \cite{Jensen:2012kj}. This raises the question on the origin of $\m$-independent contribution for further explorations.

The mass dependence of the coefficients is interesting on its own. We find small mass correction proportional to $m^2$, with correction proportional to $m^2\ln m$ vanishing. This is to be compared with mass correction found in weakly interacting massive theory at $T=0$ \cite{Gorbar:2013upa}, in which both $m^2$ and $m^2\ln m$ corrections are present. Note that in arriving at \eqref{coefficient_cse} and \eqref{coefficient_acve}, we keep only medium dependent contributions. It would be interesting to find out if the inclusion of vacuum contribution leads to logarithmic corrections as in \cite{Gorbar:2013upa}. We leave it for future work.

\begin{acknowledgments}
S.L. is supported by One Thousand Talent Program for Young Scholars and NSFC under Grant Nos 11675274 and 11735007.
\end{acknowledgments}

\appendix
\section{Stress-energy tensor for massive Dirac fermions}
The Lagrangian density for Dirac field in general relativity is
\begin{align}
  {\cal L}=\frac{i}{2}\Big(\bar{\ps}\g^ae_a^{\m}\na_{\m}\ps-(\na_{\m}\bar{\ps})e_a^{\m}\g^a\ps\Big)-m\bar{\ps}\ps,
\end{align}
with $S=\int d^4x\sqrt{\lvert g\rvert} {\cal L}$. The stress-energy tensor by definition is
\begin{align}T^{\m\n}=-\frac{2}{\sqrt{\lvert g\rvert}}\frac{\d S}{\d g_{\m\n}}.
\end{align}
To apply it, we need to use vierbeins. Then we get the Belinfante-Rosenfeld tensor
\begin{align}T_{bc}=\frac{1}{2}(T_{bc}^{(0)}+T_{cb}^{(0)}),\end{align}
where
\begin{align}T_{bc}^{(0)}=\frac{i}{2}\Big(\bar{\ps}\g_c\na_b\ps-(\na_b\bar{\ps})\g_c\ps\Big).\end{align}
There is no contribution from $\sqrt{\lvert g\rvert}$ if we use the equations of motion. Back to flat space, $\na_b$ is reduced to $\pd_b$ . Thus, we have
\begin{align}
  &T_{bc}^{(0)}=\frac{i}{2}\bar{\ps}\g_c\pd_b\ps-\frac{i}{2}(\pd_b\bar{\ps})\g_c\ps=\frac{i}{2}\bar{\ps}\g_c\pd_b\ps-\Big(\frac{i}{2}(\pd_b\bar{\ps})\g_c\ps\Big)^{\dag}=i\bar{\ps}\g_c\pd_b\ps,\no
  &T_{bc}=\frac{i}{2}(\bar{\ps}\g_b\pd_c\ps+\bar{\ps}\g_c\pd_b\ps), \qquad T^{\m\n}=\frac{i}{2}(\bar{\ps}\g^{\m}\pd^{\n}\ps+\bar{\ps}\g^{\n}\pd^{\m}\ps),\no
  &{\cal L}=i\bar{\ps}\slashed{\pd}\ps-m\bar{\ps}\ps.
\end{align}
Here we have restored the symmetric stress-energy tensor and Lagrangian density in flat space.

We can check it by the Noether's procedure by considering spacetime translations in non-gravitational theory as follows.
The canonical tensor is not symmetric for fields with spin
\begin{align}
  {\Th^\m}_\n &\equiv \frac{\pd \mathcal{L}}{\pd (\pd_\m \ps^{(i)} )}\pd_\n \ps^{(i)} - {\d^\m}_\n \mathcal{L}\no
  &=\frac{\pd \mathcal{L}}{\pd (\pd_\m \ps^{(i)} )} \pd_\n \ps^{(i)}
  \no
  &=i\bar{\ps}\g^{\m}\pd_{\n}\ps.
\end{align}
Note that the term proportional to $\cal L$ vanishes on shell.
Symmetrizing the canonical tensor, we reproduce the flat space Belinfante-Rosenfeld tensor:
\begin{align}
  T^{\m\n} \equiv \frac{1}{2}(\Th^{\m\n} + \Th^{\n\m})=\frac{i}{2}(\bar{\ps}\g^{\m}\pd^{\n}\ps+\bar{\ps}\g^{\n}\pd^{\m}\ps) \,,
\end{align}
It is both symmetric and conserved.


\begin{thebibliography}{90}

\bibitem{Vilenkin:1980fu} 
  A.~Vilenkin,
  Phys.\ Rev.\ D {\bf 22}, 3080 (1980).
  doi:10.1103/PhysRevD.22.3080



\bibitem{Erdmenger:2008rm} 
  J.~Erdmenger, M.~Haack, M.~Kaminski and A.~Yarom,
  JHEP {\bf 0901}, 055 (2009)
  doi:10.1088/1126-6708/2009/01/055
  [arXiv:0809.2488 [hep-th]].



\bibitem{Banerjee:2008th} 
  N.~Banerjee, J.~Bhattacharya, S.~Bhattacharyya, S.~Dutta, R.~Loganayagam and P.~Surowka,
  JHEP {\bf 1101}, 094 (2011)
  doi:10.1007/JHEP01(2011)094
  [arXiv:0809.2596 [hep-th]].



\bibitem{Son:2009tf} 
  D.~T.~Son and P.~Surowka,
  Phys.\ Rev.\ Lett.\  {\bf 103}, 191601 (2009)
  doi:10.1103/PhysRevLett.103.191601
  [arXiv:0906.5044 [hep-th]].



\bibitem{Neiman:2010zi} 
  Y.~Neiman and Y.~Oz,
  JHEP {\bf 1103}, 023 (2011)
  doi:10.1007/JHEP03(2011)023
  [arXiv:1011.5107 [hep-th]].



\bibitem{Landsteiner:2011cp} 
  K.~Landsteiner, E.~Megias and F.~Pena-Benitez,
  Phys.\ Rev.\ Lett.\  {\bf 107}, 021601 (2011)
  doi:10.1103/PhysRevLett.107.021601
  [arXiv:1103.5006 [hep-ph]].



\bibitem{Kharzeev:2004ey} 
  D.~Kharzeev,
  Phys.\ Lett.\ B {\bf 633}, 260 (2006)
  doi:10.1016/j.physletb.2005.11.075
  [hep-ph/0406125].



\bibitem{Kharzeev:2007tn} 
  D.~Kharzeev and A.~Zhitnitsky,
  Nucl.\ Phys.\ A {\bf 797}, 67 (2007)
  doi:10.1016/j.nuclphysa.2007.10.001
  [arXiv:0706.1026 [hep-ph]].



\bibitem{Fukushima:2008xe} 
  K.~Fukushima, D.~E.~Kharzeev and H.~J.~Warringa,
  Phys.\ Rev.\ D {\bf 78}, 074033 (2008)
  doi:10.1103/PhysRevD.78.074033
  [arXiv:0808.3382 [hep-ph]].



\bibitem{Metlitski:2005pr} 
  M.~A.~Metlitski and A.~R.~Zhitnitsky,
  Phys.\ Rev.\ D {\bf 72}, 045011 (2005)
  doi:10.1103/PhysRevD.72.045011
  [hep-ph/0505072].



\bibitem{Son:2004tq} 
  D.~T.~Son and A.~R.~Zhitnitsky,
  Phys.\ Rev.\ D {\bf 70}, 074018 (2004)
  doi:10.1103/PhysRevD.70.074018
  [hep-ph/0405216].



\bibitem{Son:2012wh} 
  D.~T.~Son and N.~Yamamoto,
  Phys.\ Rev.\ Lett.\  {\bf 109}, 181602 (2012)
  doi:10.1103/PhysRevLett.109.181602
  [arXiv:1203.2697 [cond-mat.mes-hall]].



\bibitem{Son:2012zy} 
  D.~T.~Son and N.~Yamamoto,
  Phys.\ Rev.\ D {\bf 87}, no. 8, 085016 (2013)
  doi:10.1103/PhysRevD.87.085016
  [arXiv:1210.8158 [hep-th]].



\bibitem{Stephanov:2012ki} 
  M.~A.~Stephanov and Y.~Yin,
  Phys.\ Rev.\ Lett.\  {\bf 109}, 162001 (2012)
  doi:10.1103/PhysRevLett.109.162001
  [arXiv:1207.0747 [hep-th]].



\bibitem{Pu:2010as} 
  S.~Pu, J.~h.~Gao and Q.~Wang,
  Phys.\ Rev.\ D {\bf 83}, 094017 (2011)
  doi:10.1103/PhysRevD.83.094017
  [arXiv:1008.2418 [nucl-th]].



\bibitem{Chen:2012ca} 
  J.~W.~Chen, S.~Pu, Q.~Wang and X.~N.~Wang,
  Phys.\ Rev.\ Lett.\  {\bf 110}, no. 26, 262301 (2013)
  doi:10.1103/PhysRevLett.110.262301
  [arXiv:1210.8312 [hep-th]].

\bibitem{Hidaka:2016yjf} 
  Y.~Hidaka, S.~Pu and D.~L.~Yang,
  Phys.\ Rev.\ D {\bf 95}, no. 9, 091901 (2017)
  doi:10.1103/PhysRevD.95.091901
  [arXiv:1612.04630 [hep-th]].

\bibitem{Manuel:2013zaa} 
  C.~Manuel and J.~M.~Torres-Rincon,
  Phys.\ Rev.\ D {\bf 89}, no. 9, 096002 (2014)
  doi:10.1103/PhysRevD.89.096002
  [arXiv:1312.1158 [hep-ph]].

\bibitem{Manuel:2014dza} 
  C.~Manuel and J.~M.~Torres-Rincon,
  Phys.\ Rev.\ D {\bf 90}, no. 7, 076007 (2014)
  doi:10.1103/PhysRevD.90.076007
  [arXiv:1404.6409 [hep-ph]].

\bibitem{Huang:2018wdl} 
  A.~Huang, S.~Shi, Y.~Jiang, J.~Liao and P.~Zhuang,
  Phys.\ Rev.\ D {\bf 98}, no. 3, 036010 (2018)
  doi:10.1103/PhysRevD.98.036010
  [arXiv:1801.03640 [hep-th]].

\bibitem{Kharzeev:2007jp} 
  D.~E.~Kharzeev, L.~D.~McLerran and H.~J.~Warringa,
  Nucl.\ Phys.\ A {\bf 803}, 227 (2008)
  doi:10.1016/j.nuclphysa.2008.02.298
  [arXiv:0711.0950 [hep-ph]].



\bibitem{Kharzeev:2010gr} 
  D.~E.~Kharzeev and D.~T.~Son,
  Phys.\ Rev.\ Lett.\  {\bf 106}, 062301 (2011)
  doi:10.1103/PhysRevLett.106.062301
  [arXiv:1010.0038 [hep-ph]].



\bibitem{Skokov:2009qp} 
  V.~Skokov, A.~Y.~Illarionov and V.~Toneev,
  Int.\ J.\ Mod.\ Phys.\ A {\bf 24}, 5925 (2009)
  doi:10.1142/S0217751X09047570
  [arXiv:0907.1396 [nucl-th]].



\bibitem{Deng:2016gyh} 
  W.~T.~Deng and X.~G.~Huang,
  Phys.\ Rev.\ C {\bf 93}, no. 6, 064907 (2016)
  doi:10.1103/PhysRevC.93.064907
  [arXiv:1603.06117 [nucl-th]].



\bibitem{Pang:2017bjc} 
  L.~G.~Pang, R.~H.~Fang, H.~Petersen, Q.~Wang and X.~N.~Wang,
  J.\ Phys.\ Conf.\ Ser.\  {\bf 779}, no. 1, 012069 (2017).
  doi:10.1088/1742-6596/779/1/012069



\bibitem{Voloshin:2010ut} 
  S.~A.~Voloshin,
  Phys.\ Rev.\ Lett.\  {\bf 105}, 172301 (2010)
  doi:10.1103/PhysRevLett.105.172301
  [arXiv:1006.1020 [nucl-th]].



\bibitem{Schlichting:2010qia} 
  S.~Schlichting and S.~Pratt,
  Phys.\ Rev.\ C {\bf 83}, 014913 (2011)
  doi:10.1103/PhysRevC.83.014913
  [arXiv:1009.4283 [nucl-th]].



\bibitem{Bzdak:2009fc} 
  A.~Bzdak, V.~Koch and J.~Liao,
  Phys.\ Rev.\ C {\bf 81}, 031901 (2010)
  doi:10.1103/PhysRevC.81.031901
  [arXiv:0912.5050 [nucl-th]].



\bibitem{Wang:2009kd} 
  F.~Wang,
  Phys.\ Rev.\ C {\bf 81}, 064902 (2010)
  doi:10.1103/PhysRevC.81.064902
  [arXiv:0911.1482 [nucl-ex]].



\bibitem{Wang:2012qs} 
  G.~Wang [STAR Collaboration],
  Nucl.\ Phys.\ A {\bf 904-905}, 248c (2013)
  doi:10.1016/j.nuclphysa.2013.01.069
  [arXiv:1210.5498 [nucl-ex]].



\bibitem{Bloczynski:2013mca} 
  J.~Bloczynski, X.~G.~Huang, X.~Zhang and J.~Liao,
  Nucl.\ Phys.\ A {\bf 939}, 85 (2015)
  doi:10.1016/j.nuclphysa.2015.03.012
  [arXiv:1311.5451 [nucl-th]].



\bibitem{Deng:2016knn} 
  W.~T.~Deng, X.~G.~Huang, G.~L.~Ma and G.~Wang,
  Phys.\ Rev.\ C {\bf 94}, 041901 (2016)
  doi:10.1103/PhysRevC.94.041901
  [arXiv:1607.04697 [nucl-th]].



\bibitem{Wen:2016zic} 
  F.~Wen, J.~Bryon, L.~Wen and G.~Wang,
  Chin.\ Phys.\ C {\bf 42}, no. 1, 014001 (2018)
  doi:10.1088/1674-1137/42/1/014001
  [arXiv:1608.03205 [nucl-th]].



\bibitem{Xu:2017qfs} 
  H.~j.~Xu, J.~Zhao, X.~Wang, H.~Li, Z.~W.~Lin, C.~Shen and F.~Wang,
  Chin.\ Phys.\ C {\bf 42}, no. 8, 084103 (2018)
  doi:10.1088/1674-1137/42/8/084103
  [arXiv:1710.07265 [nucl-th]].



\bibitem{Kharzeev:2015znc} 
  D.~E.~Kharzeev, J.~Liao, S.~A.~Voloshin and G.~Wang,
  Prog.\ Part.\ Nucl.\ Phys.\  {\bf 88}, 1 (2016)
  doi:10.1016/j.ppnp.2016.01.001
  [arXiv:1511.04050 [hep-ph]].



\bibitem{Huang:2015oca} 
  X.~G.~Huang,
  Rept.\ Prog.\ Phys.\  {\bf 79}, no. 7, 076302 (2016)
  doi:10.1088/0034-4885/79/7/076302
  [arXiv:1509.04073 [nucl-th]].



\bibitem{Liao:2014ava} 
  J.~Liao,
  Pramana {\bf 84}, no. 5, 901 (2015)
  doi:10.1007/s12043-015-0984-x
  [arXiv:1401.2500 [hep-ph]].



\bibitem{Lin:2018nxj} 
  S.~Lin, L.~Yan and G.~R.~Liang,
  Phys.\ Rev.\ C {\bf 98}, no. 1, 014903 (2018)
  doi:10.1103/PhysRevC.98.014903
  [arXiv:1802.04941 [nucl-th]].



\bibitem{Gorbar:2013upa} 
  E.~V.~Gorbar, V.~A.~Miransky, I.~A.~Shovkovy and X.~Wang,
  Phys.\ Rev.\ D {\bf 88}, no. 2, 025025 (2013)
  doi:10.1103/PhysRevD.88.025025
  [arXiv:1304.4606 [hep-ph]].



\bibitem{Guo:2016dnm} 
  E.~d.~Guo and S.~Lin,
  JHEP {\bf 1701}, 111 (2017)
  doi:10.1007/JHEP01(2017)111
  [arXiv:1610.05886 [hep-th]].



\bibitem{Hou:2012xg} 
  D.~F.~Hou, H.~Liu and H.~c.~Ren,
  Phys.\ Rev.\ D {\bf 86}, 121703 (2012)
  doi:10.1103/PhysRevD.86.121703
  [arXiv:1210.0969 [hep-th]].



\bibitem{Golkar:2012kb} 
  S.~Golkar and D.~T.~Son,
  JHEP {\bf 1502}, 169 (2015)
  doi:10.1007/JHEP02(2015)169
  [arXiv:1207.5806 [hep-th]].



\bibitem{Fang:2016uds} 
  R.~h.~Fang, J.~y.~Pang, Q.~Wang and X.~n.~Wang,
  Phys.\ Rev.\ D {\bf 95}, no. 1, 014032 (2017)
  doi:10.1103/PhysRevD.95.014032
  [arXiv:1611.04670 [nucl-th]].



\bibitem{Amado:2011zx} 
  I.~Amado, K.~Landsteiner and F.~Pena-Benitez,
  JHEP {\bf 1105}, 081 (2011)
  doi:10.1007/JHEP05(2011)081
  [arXiv:1102.4577 [hep-th]].

\bibitem{Jensen:2012kj} 
  K.~Jensen, R.~Loganayagam and A.~Yarom,
  JHEP {\bf 1302}, 088 (2013)
  doi:10.1007/JHEP02(2013)088
  [arXiv:1207.5824 [hep-th]].

  
\end{thebibliography}
\end{document}